\documentclass[aps,onecolumn,showpacs,preprintnumbers,amsmath,amssymb,superscriptaddress]{revtex4-2}
\usepackage{graphicx}
\usepackage{dcolumn}
\usepackage{bm}
\usepackage{color}
\usepackage{amsmath}
\usepackage{amssymb}
\usepackage{latexsym}
\usepackage{multirow}
\usepackage{xcolor}
\usepackage{hyperref}
\usepackage[capitalise]{cleveref}
\usepackage{mathtools}
\usepackage[utf8]{inputenc}
\usepackage[T1]{fontenc}
\usepackage[super]{nth}
\usepackage{lmodern}
\usepackage{tabularx}
\usepackage{gensymb}
\usepackage{upgreek}

\begin{document}
\title{Mechanical response of a thick poroelastic gel in contactless colloidal-probe rheology}
\author{Caroline Kopecz-Muller}
\affiliation{Gulliver, CNRS UMR 7083, ESPCI Paris, Univ. PSL, 75005 Paris, France.}
\affiliation{Institut Pierre Gilles de Gennes, ESPCI Paris, Univ. PSL, 75005 Paris, France.}
\affiliation{Univ. Bordeaux, CNRS, LOMA, UMR 5798, F-33400, Talence, France.}
\author{Vincent Bertin}
\email{v.l.bertin@utwente.nl}
\affiliation{Univ. Twente, Faculty Sciences and Technology, Phys. of Fluids, 7500AE Enschede, Netherlands.}
\author{Elie Rapha\"{e}l}
\affiliation{Gulliver, CNRS UMR 7083, ESPCI Paris, Univ. PSL, 75005 Paris}
\author{Joshua D. McGraw}\email{joshua.mcgraw@cnrs.fr}
\affiliation{Gulliver, CNRS UMR 7083, ESPCI Paris, Univ. PSL, 75005 Paris}
\affiliation{Institut Pierre Gilles de Gennes, ESPCI Paris, Univ. PSL, 75005 Paris, France.}
\author{Thomas Salez}\email{thomas.salez@cnrs.fr}
\affiliation{Univ. Bordeaux, CNRS, LOMA, UMR 5798, F-33400, Talence, France.}
\begin{abstract}
When a rigid object approaches a soft material in a viscous fluid, hydrodynamic stresses arise in the lubricated contact region and deform the soft material. The elastic deformation modifies in turn the flow, hence generating a soft-lubrication coupling. Moreover, soft elastomers and gels are often porous. These materials may be filled with solvent or uncrosslinked polymer chains, and might be permeable to the surrounding fluid, which complexifies further the description. Here, we derive the point-force response of a semi-infinite and permeable poroelastic substrate. Then, we use this fundamental solution in order to address the specific poroelastic lubrication coupling associated with contactless colloidal-probe methods. In particular, we derive the conservative and dissipative components of the force associated with the oscillating vertical motion of a sphere close to the poroelastic substrate. Our results may be relevant for dynamic surface force apparatus and contactless colloidal-probe atomic force microscopy experiments on soft, living and/or fragile materials, such as swollen hydrogels and biological membranes.
\end{abstract}
\maketitle

\section{Introduction}
Modern synthesis techniques and observation methods at small scales have fostered an intense research activity on soft materials and their interfaces~\cite{Andreotti_2016}. Many of these materials are elastomers and gels, deform considerably under small stress, and are commonly found in biological and technological systems. Stimuli-responsive hydrogels, for example, are candidate materials in novel devices for the detection of diseases by isolation of cells~\cite{deramo_2018}. Elastomers and gels are typically made of synthetic or natural polymers that are crossed-linked to form a network. When immersed in a favourable solvent, the polymer matrix swells as described by Flory-Huggins theory~\cite{Flory_1953}. For such gels, the material may eventually contain a large proportion of solvent molecules with a volume increase up to a factor five, for example~\cite{Li_Tran_2015}, or more. 

Solvent molecules inside a gel can diffuse through the inter-chain regions of the polymer matrix. Therefore, when a load is applied to a gel, material deformations are time-dependent. This dynamic response arises since the chains are restrained from unbounded displacements by elasticity while the solvent motion implies viscous dissipation. The latter mechanism is called poroelasticity, which has been extensively used to describe the mechanical properties of gels. The first poroelastic theory was introduced by Biot to model the consolidation of soils~\cite{Biot_1941}. Since then, additional features have been added such as nonlinear elasticity~\cite{hong2008theory,bouklas2012swelling}, viscoelasticity~\cite{hu2012viscoelasticity}, or surface stresses~\cite{Zhao_2018, liu2020coupled,ang2020effect}, and instabilities have been considered~\cite{Dervaux_2012}. Notably, the mechanical response of a gel depends on the interactions between the gel and its environment. If the gel is indented by a rigid object, as in the emblematic example of contact mechanics~\cite{johnson_1987,hui2006contact,Hu_2010,hu2011indentation,Delavoipiere_2016,Degen_2020}, solvent molecules do not flow across the interface between the gel and the indenter, which is thus impermeable. In contrast, if the gel is immersed in its own solvent, then the solvent molecules can be transported across the gel-solvent interface, which is thus permeable. 

In addition to the permeability boundary condition above, when a rigid object moves in a viscous fluid near a soft surface, it generates hydrodynamic stresses~\cite{Reynolds_1886} that, despite the lack of direct contact, may deform the soft surface. In turn, the deformation of the soft surface modifies the flow which generates so-called soft-lubrication couplings, that are at the heart of the recent development of gentle, contactless rheological methods for soft materials~\cite{Vakarelski_2010,Leroy_Charlaix_2011,Leroy_Charlaix_2011,dowson2014elasto,Kaveh_2014,Wang_2015,Wang_2017_SM,Wang_2017_COCI,Wang_2018,Karan_2020}. These methods have been employed to measure the rheology of diverse surfaces such as elastomers~\cite{Leroy_2012}, gels~\cite{guan2017noncontactbis, zhang2022contactless}, glasses~\cite{Villey_2013}, living cells~\cite{guan2017noncontact}, and liquid-air interfaces with impurities~\cite{Wang_Maali_2018,Bertin_bubble_2021}. They typically involve colloidal-probe atomic force microscopy~\cite{Cappella_Dietler_1999, Butt_2005}, dynamical surface force apparati~\cite{Israel_1976, Israel_2010, Kristiansen_2019}, or tuning-fork microscopes~\cite{laine2019micromegascope}. The underlying principle involves driving a spherical probe near a soft surface of interest (see Fig.~\ref{fig:schema}(b)), and to combine the measured force and the soft-lubrication model in order to  infer the material rheology.

Another interesting aspect of soft-lubrication couplings is the emergence of inertial-like forces at zero Reynolds number, such as the lift force for transverse driving ~\cite{Sekimoto_1993,Beaucourt_2004,Skotheim_Maha_2005,Urzay_2007,Snoeijer_2013,Salez_Maha_2015,Vialar_2019,Zhang_2020,Bertin_lift_2022,Bureau_2022}. As a direct consequence, the effective friction between two objects in respective sliding motion can be strongly reduced~\cite{Bouchet_2015,Saintyves_2016,Davies_2018,Rallabandi_2018}, as compared to the classical rigid lubrication case. This might have important practical implications, in physiology for instance, since the friction between bones in mammalian joints~\cite{Hou_1992,Jahn_Klein_2018} may be strongly reduced through the presence of poroelastic cartilages between the solid bones and the synovial lubricant. In the same way, the wet contact between the eyelid and the eyeball is complemented by a stratification of polymer-like and gel-like layers, that may offer better sliding when blinking~\cite{Cher_2008}. 

Despite the above interest in soft materials and their rheology measured from contactless methods, it is interesting to realize that only simple linear elastic-like constitutive responses have been addressed in the context of soft-lubrication theory. While the effects of viscoelasticity have been recently investigated in more detail~\cite{pandey2016lubrication,Kargar_2021,hui2021friction}, the effects of  poroelasticity remain scarcely and partially addressed~\cite{Delavoipiere_2018,Ciapa_2020, cuccia2020pore}, and certainly at a too basic level to address the more complex and subtle responses of nonlinear functionalized materials~\cite{Alaa_2022}. Early works on purely porous substrates suggest an equivalent description involving effective slippage at the interface, with either the slip length considered to be on the order of the pore size~\cite{Beavers_1967}, or a full slip boundary condition~\cite{Meeker_2004}, while more recent work suggests a key role of the permeability boundary conditions~\cite{Knox_2017}. 

In this article, in view of the identified gap in the literature noted above, we derive a model  to characterize the mechanical response of thick poroelastic gels in the framework of contactless colloidal-probe rheological methods. Since we are interested in describing gels in contact with a reservoir of solvent, a full permeability boundary condition at the gel-solvent interface is considered. Our focus contrasts with a previous study in the impermeable case~\cite{Zhao_2018, PVdV_2022}, that is more relevant to methods involving direct solid contact. In the first part, we obtain the surface deformation of a semi-infinite, permeable, poroelastic layer under the action of an arbitrary pressure field. We use the formalism of Green's functions in axisymmetric conditions, as it was done for purely elastic media in the context of soft lubrication~\cite{Hughes_1979,Kaveh_2014,Wang_2015,Karan_2020, Essink2020}, or for a poroelastic but impermeable soft substrate~\cite{Zhao_2018}. In the second part, we apply this formalism
to the canonical situation for contactless colloidal-probe rheology, namely the soft-lubricated motion of a sphere oscillating vertically near a poroelastic gel, and we characterize the substrate deformation and resulting normal force in detail.

\section{Mechanical response to an external pressure field}
\label{sec:sudden-response}
\subsection{Linear poroelastic theory}
\begin{figure}[t!]
	\includegraphics[width=16cm]{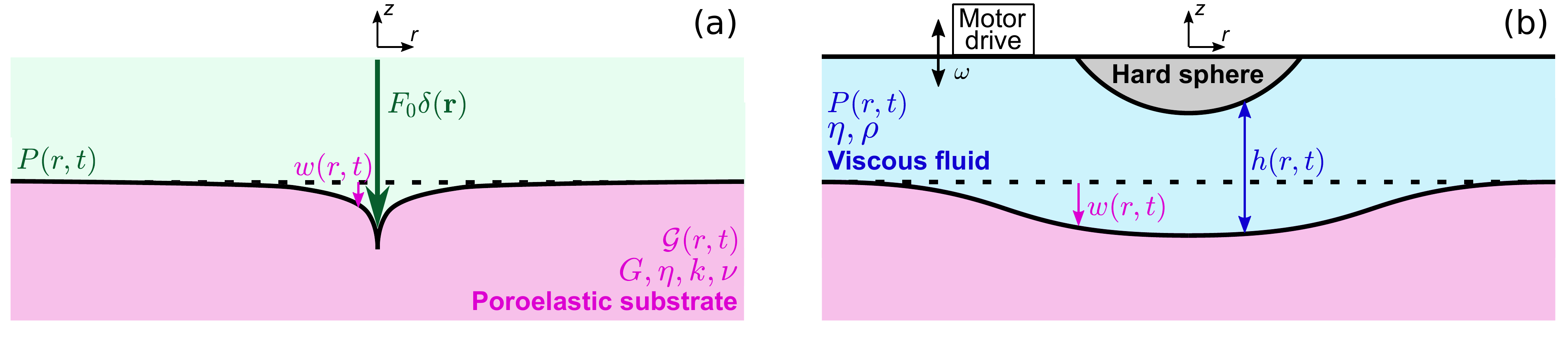}
	\caption{ \textbf{Semi-infinite permeable poroelastic medium deformed by two different external axisymmetric pressure fields.} \textit{(a) A point-force pressure field $P(r,t) = F_0\delta(\mathbf{r})$ is suddenly applied at $t=0$ on the poroelastic substrate, and generates a surface deformation $w(r,t)$. The latter is directly related to the Green's function $\mathcal{G}(r,t)$. We denote $G$, $\nu$, and $k$, the effective shear elastic modulus, effective Poisson ratio, and porosity of the substrate, respectively, as well as $\eta$ the viscosity of the solvent flowing in the porous material. (b) In contactless colloidal-probe rheological methods, a micrometric sphere is oscillating at angular frequency $\omega$ normally to the substrate in a liquid (identical to the solvent in the gel here) of dynamic shear viscosity $\eta$ and density $\rho$. The hydrodynamic lubrication pressure field $P(r,t)$ generated by the associated flow deforms the gel surface, leading to a deformed liquid gap profile $h(r,t)$.}}
	\label{fig:schema}
\end{figure}

The considered system is shown in Fig.~\ref{fig:schema}(a). It consists of a gel that occupies the half-space defined by $z\leq 0$. We suppose that the mechanics of the gel is described by the linear poroelastic theory. As mentioned in the introduction, this model was first established by Biot~\cite{Biot_1941}, and was adapted to model the migration of solvent in elastomeric gels~\cite{hui2006contact,Zhao_2018,PVdV_2022}. We take as a reference state a swollen gel, with a homogeneous solvent concentration $c_0$, and where the chemical potential of the solvent inside the gel is $\mu_0$. The elastic deformation of the gel is characterized by the strain tensor $\mathbf{\epsilon}$. The latter is defined as the symmetric part of the displacement field gradient tensor, as:
\begin{equation}
	\mathbf{\epsilon} = \frac{1}{2}\left[ \mathbf{\nabla}\mathbf{u} + (\mathbf{\nabla}\mathbf{u})^T\right],
	\label{eq:strain-def}
\end{equation}
where $\mathbf{u}$ denotes the displacement field with respect to the reference state. The solvent mass being conserved, the concentration $c$ satisfies the continuity equation:
\begin{equation}
	\displaystyle \frac{\partial c}{\partial t} + \mathbf{\nabla}\cdot \mathbf{J}=0,
	\label{eq:adv}
\end{equation}
where the flux of solvent inside the gel is denoted $\mathbf{J}$. The linear poroelastic theory assumes that the solvent flow is driven by the gradient of solvent chemical potential $\mu$, through the Darcy law:
\begin{equation}
        \mathbf{J} = -\Big(\frac{k}{\eta \Omega^2} \Big) \mathbf{\nabla} \mu.
        \label{eq:darcy}
\end{equation}
Here, $\eta$ and $\Omega$ are the viscosity and molecular volume of the solvent, respectively, and $k$ is the permeability that is on the order of the pore surface area of the swollen polymeric network -- but with a prefactor depending on the specific network architecture. The solvent and the polymer network are both supposed to be incompressible. As a consequence, the local variations of volume of the polymer network are due to the local changes in solvent concentration, which sets the condition:
\begin{equation}
	\text{Tr}(\mathbf{\epsilon})=\mathbf{\nabla}\cdot \mathbf{u}=(c-c_0)\Omega,
	\label{eq:div-def}
\end{equation}
where Tr is the trace. As discussed in Ref.~\cite{hu2011indentation}, we expect that the free energy density $\mathcal{U}$ of the gel is a function of the strain tensor and the concentration field. The work done on a gel element is given by $\delta \mathcal{U} = \sigma_{ij}\delta \epsilon_{ij} + (\mu -\mu_0)\delta c$, where $\sigma$ is the mechanical stress tensor. Nevertheless, because of the incompressibility condition in Eq.~\eqref{eq:div-def}, the solvent concentration is no longer an independent variable, and the free energy density only depends on strain. The latter is supposed to follow the standard linear-elastic energy density, \textit{i.e.} $\mathcal{U}=G\left[\mathbf{\epsilon}:\mathbf{\epsilon} + \frac{\nu}{1-2\nu} \mathrm{Tr}^2(\mathbf{\epsilon})\right]$, where $G$ and $\nu$ are the effective shear elastic modulus and Poisson ratio, respectively. The stress tensor is then given by~\cite{hu2011indentation}:
\begin{equation}
    \mathbf{\sigma}=2G\Big[\mathbf{\epsilon} +\frac{\nu}{1-2\nu}\text{Tr}(\mathbf{\epsilon})\mathbf{I}\Big]-\frac{\mu-\mu_0}{\Omega}\mathbf{I},
    \label{eq:stress-strain}
\end{equation}
where $\mathbf{I}$ is the identity tensor. The difference in chemical potentials per molecular volume appears as a hydrostatic pressure, often called \textit{pore pressure}, and is obtained by enforcing the incompressibility condition with a Lagrange multiplier. In the absence of body force, the mechanical equilibrium is expressed by Navier's closure equation: 
\begin{equation}
    \mathbf{\nabla}\cdot\mathbf{\sigma}=\mathbf{0}.
    \label{eq:navier}
\end{equation}
Combining the two last equations leads to:
\begin{equation}
	\label{eq:navier-fields}
G \Omega\left(\mathbf{\nabla}^2 \mathbf{u}+\frac{\Omega}{1-2 \nu} \mathbf{\nabla} (c-c_0)\right)=\mathbf{\nabla} (\mu-\mu_0)  .
\end{equation}
Invoking Eq.~\eqref{eq:darcy}, Eqs. \eqref{eq:adv},\eqref{eq:div-def}, and \eqref{eq:navier-fields} form a closed system of five equations for the five fields $\mu$, $c$, and the three components of $\mathbf{u}$. Combining the latter equations reduces the problem to a set of two coupled equations on the concentration field $c$ and chemical potential $\mu$, as:
\begin{subequations}
\label{eq:poroelasticreduced}
\begin{equation}
	\label{eq:c_mu_coupled}
	\displaystyle \mathbf{\nabla}^2\left[(\mu-\mu_0)-2G\Omega^2 \frac{1-\nu}{1-2\nu}(c-c_0) \right]=0, 
\end{equation}
\begin{equation}
	\label{eq:diffusion}
	\displaystyle \frac{\partial c}{\partial t} = \mathcal{D}_\mathrm{pe}\mathbf{\nabla}^2 c,
\end{equation}
\end{subequations}   
where we have introduced an effective, poroelastic diffusion coefficient
\begin{equation}
	\mathcal{D}_\mathrm{pe}= \frac{2(1-\nu)}{1-2\nu}\frac{Gk}{\eta}.
\label{eq:EffectiveDiffusion}
\end{equation}
Equation~\ref{eq:c_mu_coupled} couples the chemical potential with the concentration, as the flow of solvent is driven by gradients of chemical potential (or equivalently, gradients of pore pressure). Equation~\ref{eq:diffusion} describes the diffusion of solvent trough the porous matrix, with  $\mathcal{D}_\mathrm{pe}$ of Eq.~\ref{eq:EffectiveDiffusion} constructed using macroscopic material parameters. We note lastly, however, that even while $\mathcal{D}_\mathrm{pe}$ is constructed from these macroscopic parameters, one can recover a molecular-scale diffusion coefficient. To make the correspondence, we use the Stokes-Einstein relation for molecular diffusion, $\mathcal{D}_\upmu \sim k_{\textrm{B}}T/(\eta a)$, with $kT$ and $a\approx\Omega^{1/3}$ the thermal energy and monomer size, respectively. Estimating furthermore the typical polymeric modulus $G \sim k_{\textrm{B}}T/(Na^3)$, and the permeability $k\sim Na^2$, where $N$ is a typical number of monomers between crosslinks in the network, we find $\mathcal{D}_\mathrm{pe} \sim \mathcal{D}_\upmu$ upon substitution into Eq.~\ref{eq:EffectiveDiffusion}. 

\subsection{Point-force driving}
\label{sec:boundary_conditions}
We now derive the response of the gel to a spatially delta-distributed force density applied to the surface of the gel. Prior to the application of such a force, \emph{i.e.} for times $t \leq 0$, we suppose that the gel is in the (swollen) reference state with strain- and stress-free conditions. For $t\geq0$, a point-force pressure source of magnitude $F_0$ is suddenly applied on the surface. This forcing drives a deformation of the gel surface, and solvent flow within the polymer matrix. At the interface (\textit{i.e.} $z=0$ in the reference state), the stress boundary condition is therefore given by:
\begin{equation}
\mathbf{\sigma} \cdot \mathbf{e}_z = -F_0 \delta(\mathbf{r})H(t)  \mathbf{e}_z,
\label{eq:DELTA}
\end{equation}
where $H(t)$ denotes the Heaviside step function and $\delta(\bold{r})$ the Dirac distribution.

In the limit $z\rightarrow -\infty$, the stress and strain fields vanish and the solvent concentration field reaches its reference equilibrium value $c_0$. At infinitesimally small times after the point force has been applied, the solvent did not have time to flow, so that the solvent concentration is the same as the one at $t<0$, \textit{i.e.}:
\begin{equation}
	c(r,z,t=0) = c_0.
\end{equation} 
We suppose that the gel is in contact with a reservoir of solvent molecules, which sets the surface chemical potential to the reference equilibrium value $\mu_0$. Such a permeability condition allows for solvent exchange between the gel and the outer reservoir, and is relevant to situations where the gel is immersed in a liquid phase (\textit{e.g.} its own solvent) with some affinity between the two. Thus, the boundary condition on the chemical potential at the interface reads:
\begin{equation}
    \mu(r,z=0,t) = \mu_0. \\
\label{eq:BC_chempot}
\end{equation}

\subsubsection{Resolution}
To determine the surface deformation $w(r,t)$ associated with the pressure source of Eq.~\eqref{eq:DELTA} (see Fig.~\ref{fig:schema}(a)), we follow the method introduced by McNamme \& Gibson~\cite{MacNamee_Gibson_1,MacNamee_Gibson_2,Gibson_1970,liu2020coupled}. The key ingredient of that method is the introduction of two displacement potentials $A(r,z,t)$ and $B(r,z,t)$, defined by:
\begin{subequations}
	\begin{equation}
	\label{eq:radial_displacement}
		\displaystyle u_r = z \frac{\partial A}{\partial r}+ \frac{\partial B}{\partial r}, \\
	\end{equation}
	\begin{equation}
		\label{eq:normal_displacement}
		\displaystyle u_z = z \frac{\partial A}{\partial z} - A + \frac{\partial B}{\partial z}, \\
	\end{equation}
	\label{eq:disp}
	\end{subequations}
and that satisfy the following equations:
	\begin{subequations}
	\begin{equation}
		\displaystyle \mathbf{\nabla}^2 A = 0,
	\end{equation}
	\begin{equation}
		\nabla^2 B=\Omega(c-c_0),
			\label{eq:eq_potential}
	\end{equation}
	\begin{equation}
	2G\Omega\displaystyle \frac{\partial A}{\partial z} = (\mu-\mu_0)-2G\Omega^2 \frac{1-\nu}{1-2\nu}(c-c_0),
		\end{equation}
	\begin{equation}
		\displaystyle \frac{\partial \mathbf{\nabla}^2 B}{\partial t} = \mathcal{D}_\mathrm{pe}\mathbf{\nabla}^4 B.
	\end{equation}
	\label{eq:potential}
\end{subequations}
Using Eq.~\eqref{eq:stress-strain},~\eqref{eq:disp} and~\eqref{eq:potential}, the components of the stress tensor can be expressed as:
\begin{subequations}
	\label{eq:stress_potential}
	\begin{equation}
		\displaystyle \sigma_{rr} = 2G\left( z \frac{\partial^2 A}{\partial r^2} - \frac{\partial A}{\partial z}+ \frac{\partial^2 B}{\partial r^2} - \Delta B  \right), \\
	\end{equation}
	\begin{equation}
		\displaystyle \sigma_{zz} =  2G\left( z \frac{\partial^2 A}{\partial z^2} - \frac{\partial A}{\partial z}+\frac{\partial^2 B}{\partial z^2} - \Delta B  \right), \\
	\end{equation}
	\begin{equation}
		\displaystyle\sigma_{rz} = 2G \left(\frac{\partial^2 B}{\partial r \partial z} + z \frac{\partial^2 A}{\partial z \partial r} \right). \\
	\end{equation}
\end{subequations}
We note that azimuthal stresses and displacements have not been considered here, given the axisymmetry of the problem. 

To solve Eqs.~\ref{eq:eq_potential}, we reconsider the problem in the spectral domain. Specifically, we use the Hankel transform of zeroth order in space and the Laplace transform in time. In such a framework, a given field $X(r,t)$ is transformed into:
\begin{equation}
	\hat{X}(s,q) =\int_{0}^{\infty} \text{d}t \, e^{-q t}\int_0^\infty \text{d}r\text{ } X(r,t)  r J_0(sr),
	\label{eq:space_forward}
\end{equation}
where $J_0$ is the Bessel function of the first kind and zeroth order. The inversion formula reads:
\begin{equation}
	\displaystyle X(r,t) = \frac{1}{2\pi i}  \int_{\gamma -i \infty}^{\gamma +i\infty} \text{d}q \, e^{q t} \int_0^\infty \text{d}s\, \hat{X}(s,q)  s J_0(sr),
	\label{eq:inversion}
\end{equation}
where the inverse Laplace transform is written using the Bromwich integral. Then, expressing Eqs.~\ref{eq:eq_potential} in the spectral domain and invoking the initial condition $\nabla^2 B(r,z,0)=0$, we get the following ordinary differential equations on the transformed potentials $\hat{A}(s,z,q)$ and $\hat{B}(s,z,q)$:
\begin{eqnarray}
	\label{eq:eq_pot_HL}
		\left(\frac{\partial^2}{\partial z^2} - s^2 \right)\hat{A} &=0, \\
			\label{eq:eq_pot_HL2}
		\left(\frac{\partial^2}{\partial z^2} - s^2 -\frac{q}{\mathcal{D}_\mathrm{pe}} \right)\left(\frac{\partial^2}{\partial z^2} - s^2 \right)\hat{B} &=0.
\end{eqnarray}
The solutions to Eqs.~\eqref{eq:eq_pot_HL} and~\eqref{eq:eq_pot_HL2} that vanish at $z \rightarrow -\infty$ read:
\begin{subequations}
	\label{eq:formal_solutions}
	\begin{equation}
		\displaystyle \hat{A} = a_1 e^{sz},\\
	\end{equation}
	\begin{equation}
		\displaystyle \hat{B} = b_1 e^{sz} + b_2 e^{z\sqrt{s^2+q/\mathcal{D}_\mathrm{pe}}},\\
	\end{equation}
\end{subequations}
where $a_1, b_1, b_2$ are integration constants, that depend on the spectral variables $s$ and $q$. Expressing the stress and chemical-potential boundary conditions of Eqs.~\eqref{eq:DELTA} and~\eqref{eq:BC_chempot} in terms of the potentials, we obtain:
\begin{subequations}
	\label{eq:BC_transformed}
	\begin{equation}
		\displaystyle \hat{\sigma}_{sz}(s,z=0,q) = 0 = -2Gs\left[b_1s + b_2\sqrt{s^2 +\frac{q}{\mathcal{D}_\mathrm{pe}}}\right],  \\
	\end{equation}
	\begin{equation}
		\displaystyle\hat{\sigma}_{zz}(s,z=0,q) =-\frac{ F_0}{2\pi q} = 2G\left[-a_1s +(b_1+b_2)s^2\right],\\
	\end{equation}
	\begin{equation}
		\displaystyle \hat{\mu}(s,z=0,q)- \hat{\mu}_0= 0 =  2G\Omega\left[\frac{1-\nu}{1-2\nu}b_2\frac{q}{\mathcal{D}_\mathrm{pe}}+a_1s \right].
	\end{equation}
\end{subequations}
Solving Eqs.~\eqref{eq:BC_transformed}, we obtain $a_1$, $b_1$ and $b_2$. 

\begin{figure}[t!]
	\centering
	\includegraphics[width=16cm]{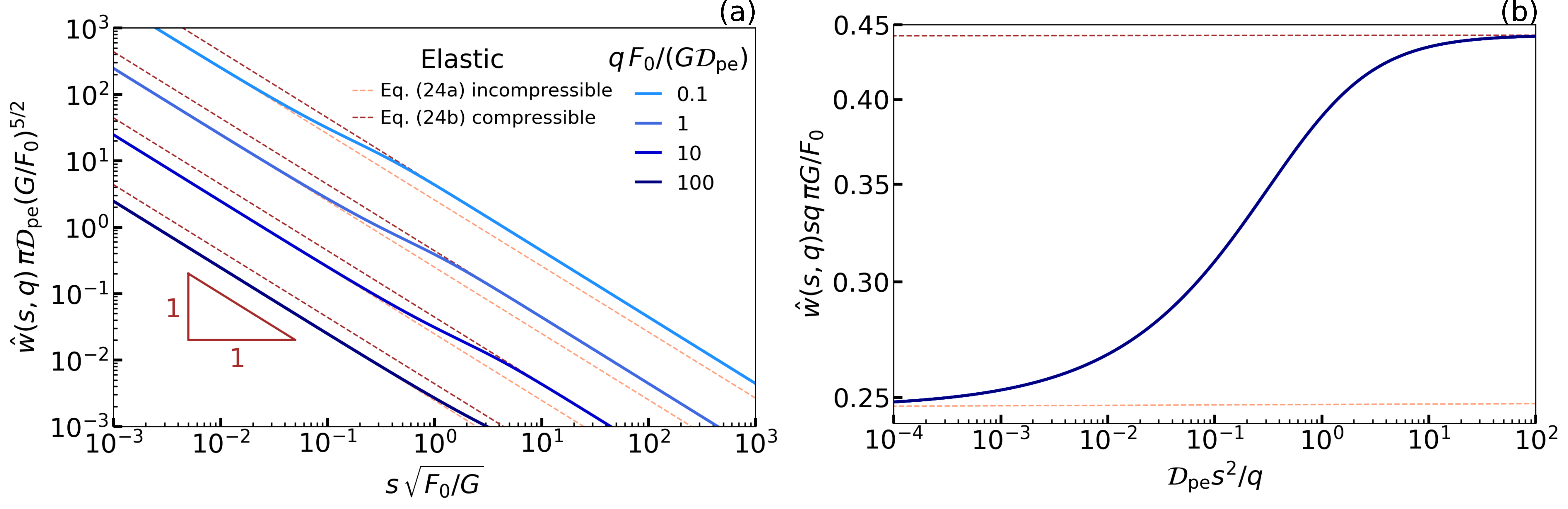}
	\caption{\textbf{Surface deformation induced by a point-force pressure source in Hankel-Laplace space.} \textit{(a) Dimensionless, reciprocal-space surface deformation $\hat{w}(s,q)$ as computed from Eq.~\eqref{eq:green}, as a function of the scaled spatial frequency, $s$, for various temporal frequencies, $q$, normalized by $\mathcal{D}_\mathrm{pe}G/F_0$, using $\nu=0.1$. The orange and red dashed lines correspond to Eqs.~\eqref{eqshort} and~\eqref{eqlong}, respectively. (b) Normalized surface deformation in reciprocal space $\hat{w}(s,q)$, as a function of the variable $\mathcal{D}_\mathrm{pe}s^2/q$, for $\nu=0.1$, as computed from Eq.~\eqref{eq:green}. The orange and red dashed lines as described in (a). }}
	\label{fig:Point-force_response_spectral}
\end{figure}

\subsubsection{Result and discussion}
Having solved the poroelastic problem Eqs.~\ref{eq:poroelasticreduced}-\ref{eq:BC_chempot} for the potentials $A,B$ defined in the previous section, we find the surface normal deformation of the gel $\hat{w}(s,q) = -\hat{u}_z(s,z=0,q)$ in reciprocal space (see Eq.~\ref{eq:space_forward}) by invoking Eq.~\eqref{eq:normal_displacement}, as:
\begin{equation}
	\displaystyle \hat{w}(s,q) =  \frac{F_0}{4\pi Gsq}\,\frac{1}{1+\Lambda\frac{\mathcal{D}_\mathrm{pe}s^2}{q}\Big(1-\sqrt{1+\frac{q}{s^2\mathcal{D}_\mathrm{pe}}}\Big)}\ ,
	\label{eq:green}
\end{equation}
where the Poisson ratio appears in the compressiblity factor 
\begin{equation}
	\Lambda = \frac{1-2\nu}{1-\nu}. 
	\label{eq:LambdaDef}
\end{equation}
We first note that if the gel is nearly incompressible, \textit{i.e.} as $\nu \rightarrow 1/2$, then $\Lambda \rightarrow 0$ and the poroelasticity does not affect the surface deformation as revealed in Eq.~\eqref{eq:green}. In such a case, the poroelastic medium responds as a purely elastic and incompressible one, at all times. Similarly, if the permeability is small, \textit{i.e.} as $k\rightarrow 0$, the diffusion constant $\mathcal{D}_\mathrm{pe}$ of the solvent vanishes, and the medium again behaves as an incompressible elastic material. In the opposite limit of large permeability, the solvent can diffuse almost instantaneously, and the stress is immediately relaxed, so that the response is one of a compressible elastic material at all times. 

In Fig.~\ref{fig:Point-force_response_spectral}(a) is shown $\displaystyle \hat{w}(s,q)$ plotted as a function of $s$ for various $q$. We choose a value $\nu=0.1$, a typical value for swollen gels and giving finite $\Lambda$. The vertical normalisation is chosen such that dimensionless values of $s$ and $q$ were used, length is normalised by $\sqrt{F_0/G}$, and time by $F_0/(G\mathcal{D}_\mathrm{pe})$. The results show parallel power-law decays in the small- and large-$s$ limits, with a larger prefactor for large $s$ (small distance). To explain this observation, we explore the temporal asymptotics of the governing Eq.~\ref{eq:green}. The initial and final value theorems can be used in the short-time and long-time limits of the surface deformation. We thus find:
\begin{subequations}
\label{asymptotereciprocal}
\begin{equation}
\label{eqshort}
	\displaystyle \hat{w}(s,t=0^+) = \lim_{q \rightarrow \infty} q\hat{w}(s,q) = \frac{F_0}{4\pi Gs}, 
\end{equation}
\begin{equation}
\label{eqlong}
	\displaystyle \hat{w}(s,t\rightarrow\infty) = \lim_{q \rightarrow 0^+} q\hat{w}(s,q) = \frac{F_0(1-\nu)}{2\pi Gs}, 
\end{equation}
\end{subequations}
leading to the deformation in real space:
\begin{subequations}
	\label{asymptotereal}
	\begin{equation}
		\label{eq:incompressible_limit}
		\displaystyle w(r,t=0^+) = \frac{F_0}{4\pi Gr} = w^\mathrm{incomp}(r), 
	\end{equation}
	\begin{equation}
		\label{eq:compressible_limit}
		\displaystyle w(r,t\rightarrow\infty)= \frac{F_0(1-\nu)}{2\pi Gr} = w^\mathrm{comp}(r).
	\end{equation}
\end{subequations}
Thus for both short and long times, we find spectral power-law decays of the surface deformation. The former expression is the point-force solution of a purely elastic, incompressible, and semi-infinite medium of shear modulus $G$, denoted $w^\mathrm{incomp}(r)$. At long times, we have the point-force solution of a purely elastic and semi-infinite medium of shear modulus $G$ and Poisson ratio $\nu$, denoted $w^\mathrm{comp}(r)$. Equations~\ref{asymptotereciprocal} are plotted using dashed lines in Fig.~\ref{fig:Point-force_response_spectral}(a). These expressions thus form a link between poroelastic and elastic materials~\cite{Skotheim_Maha_2005, Hariprasad_2012}: at large distances (small $s$), the solvent has no time to flow inside the porous matrix and the response is elastic-like, with an incompressibility condition due to the liquid fraction. At small distances (large $s$), the solvent does not flow anymore and the response recovers a steady elastic deformation, with compression (\emph{i.e.} a concentration change) as compared to the initial state. In between the two asymptotic regimes, the surface deformation smoothly changes from the short-time (incompressible) to the long-time (compressible) elastic-like behaviours, as shown with blue lines and using finite $q$ in Fig.~\ref{fig:Point-force_response_spectral}(a).

To connect the asymptotic inverse space and time responses, we note in Eq.~\eqref{eq:green} that at fixed $\Lambda$, a natural variable is an inverse, dimensionless diffusive one $\mathcal{D}_\mathrm{pe}s^2/q$. This is expected since the solvent concentration follows a diffusive-like law with a diffusion constant $\mathcal{D}_\mathrm{pe}$. In Fig.~\ref{fig:Point-force_response_spectral}(b), we thus plot the normalized surface deformation in reciprocal space, as a function of the normalized diffusive variable, having fixed the Poisson ratio and for the same $q$ noted in Fig.~\ref{fig:Point-force_response_spectral}(b). Given the normalisation and the form of Eq.~\ref{eq:green}, we find that a single curve describes the response in reciprocal space. Interpreting the response physically, we note that when the gel starts to be indented, it first exhibits an incompressible elastic-like response, as discussed above. Later, at a given time $t$, the solvent and stress have typically diffused over a radial distance $r_{\textrm{c}}\sim\sqrt{\mathcal{D}_\mathrm{pe} t}$, giving a self-similar curve in reciprocal space. 

\begin{figure}[t!]
	\centering
	\includegraphics[width= 16cm]{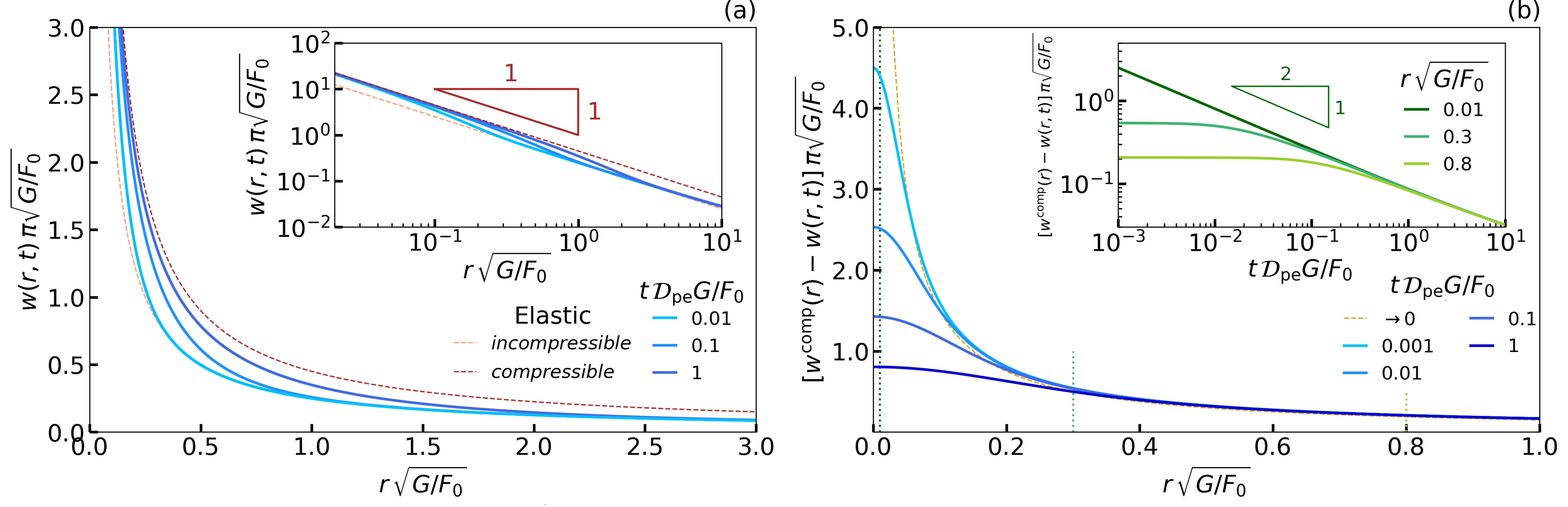}
	\caption{\textbf{Surface deformation induced by a point-force pressure source.} \textit{(a) Normalized surface deformation as a function of the radial coordinate, for the times noted in the legend, as computed from the inverse transform of Eq.~\eqref{eq:green} and using $\nu=0.1$. The orange and red dashed lines correspond to the asymptotic limits in Eqs.~\eqref{asymptotereal}. The inset shows the same data on logarithmic scales. (b) Difference $w^{\textrm{comp}}(r)-w(r,t)$ (normalized) between the surface deformation of a purely elastic compressible material (see Eq.~\eqref{eq:compressible_limit}) and the one of the poroelastic material, as a function of normalized radial coordinate, for the dimensionless times noted in the legend. The inset shows the same data, but as a function of time and for various radial positions (corresponding to the vertical dashed green lines in the main panel), using logarithmic scales. The $-1/2$ exponent of the asymptotic power-law behaviour is discussed in Appendix~\ref{APP:long-time-power}.}}
	\label{fig:Point-force_response-real_space}
\end{figure}

To have a direct view on the spatial and temporal relaxations described above for reciprocal space, the inverse Laplace transform of Eq.~\ref{eq:green} was numerically computed using the Talbot algorithm~\cite{Abate_Whitt_2006}. The inverse Hankel transform was computed with Riemann summation over a finite spectral domain. Residual numerical oscillations were smoothed using a Savitzky-Golay filter of order 3 on a window of 9 points over the total 200\,000 used in the linear discretisation of the $r$, $s$ axes. The results are presented in Fig.~\ref{fig:Point-force_response-real_space}(a), where the deformation in real space is plotted as a function of the radial coordinate for various times. For $r< r_{\textrm{c}}$ noted in the previous paragraph, the gel state has essentially relaxed and the response is compressible (red dashed line), while for $r> r_{\textrm{c}}$ the state and response are not modified with respect to the initial, incompressible elastic ones (orange dashed line). The transition between compressible and incompressible deformations are also elucidated in the logarithmic representation of the data shown in the inset, where the short- and long-time asymptotic relaxations are shown. 

In Fig.~\ref{fig:Point-force_response-real_space}(b), we quantitatively show the gel's relaxation to its final state, plotting the difference of the the data in (a) with that of the asymptotic late-time limit in Eq.~\eqref{eq:compressible_limit} as a function of the radial coordinate. A continuous decay toward the late-time value is observed for all radii. Taking a few examples, we show in the inset of Fig.~\ref{fig:Point-force_response-real_space}(b) the temporal decay toward the final state for the three radii noted by vertical dashed lines in the main part of the figure. For early times, we note a plateau at a value $F_0(1-2\nu)/(4\pi G)$, corresponding to the difference between the two asymptotic limits in Eq.~\eqref{asymptotereal}. Remarkably, a temporal power law with exponent $-1/2$, characteristic of a diffusive process, is reached for all the radii at long times. In Appendix~\ref{APP:long-time-power}, the late-time $t^{-1/2}$ asymptotic power-law decay is demonstrated by expanding Eq.~\ref{eq:green} at small $q$ and transforming to real space. The intercept between the asymptotic decay law and the initial plateau value indicates its typical duration time, that scales with the diffusion time $r^2/\mathcal{D}_\mathrm{pe}$.

Lastly, we note that in Appendix~\ref{app:impermeable_comp}, we compare the results of the present permeable description to the case of an impermeable surface. For the impermeable case, the solvent flux vanishes at the interface. This alternative boundary condition is relevant when the gel is not in contact with its own liquid solvent. Such a situation arises when a gel is indented by a rigid object~\cite{johnson_1987,hui2006contact,Hu_2010,hu2011indentation,Delavoipiere_2016,Degen_2020}, as well as in some configurations of soft wetting~\cite{Zhao_2018,Xu_2020}. The surface deformations are found to adopt qualitatively similar shapes in the permeable and impermeable cases. However, the respective behaviours quantitatively differ, and the stress relaxation is in particular faster in the permeable case, due to the allowed exchange of solvent with the outer reservoir.

\subsection{Solution for an arbitrary pressure field}
\label{sec:generalization}
In real systems, gels are indented with probes that have finite size~\cite{Degen_2020, Hu_2010, hu2011indentation, Delavoipiere_2016}. In these cases, the external load is not a point force, and the outer pressure field has a finite spatial extent. Additionally, the outer pressure field may exhibit temporal variations. Since the above model only involves linear operators, we can apply the superposition principle. Henceforth, the surface deformation generated by an arbitrary time-dependent and space-dependent pressure field $p(\mathbf{r},t)$, is given by the convolution:
\begin{equation}
	\displaystyle w(\mathbf{r},t) = \int_{-\infty}^{t} \text{d}t' \int_{\mathbb{R}^2}\mathrm{d}^2\mathbf{r}'\, \mathcal{G}(\lvert\mathbf{r}-\mathbf{r}'\rvert,t-t')\,p(\mathbf{r}',t'),
	\label{eq:response}
\end{equation}
where $\mathcal{G}$ is the Green's function of the problem, which is the surface deformation induced by a point force $\delta(\mathbf{r})\delta(t)$. The latter is directly related to Eq.~\eqref{eq:green}, through:
\begin{equation}
	\displaystyle \hat{\mathcal{G}}(s,q) = \frac{1}{4\pi Gs}\,\frac{1}{1+\Lambda\frac{\mathcal{D}_\mathrm{pe}s^2}{q}\Big(1-\sqrt{1+\frac{q}{s^2\mathcal{D}_\mathrm{pe}}}\Big)}\ ,
	\label{eq:response2}
\end{equation}
and the inverse transform:
\begin{equation}
	\displaystyle \mathcal{G}(r,t) = \frac{1}{2i\pi} \int_{\gamma -i \infty}^{\gamma +i\infty} \text{d}q\, e^{q t}\int_0^\infty \text{d}s\text{ } \hat{ \mathcal{G}}(s,q) s J_0(sr) .
	\label{eq:response3}
\end{equation}

\section{Application to contactless colloidal-probe rheology}
In this section, we apply the general results of the previous one for a specific outer pressure field that is relevant to contactless colloidal-probe rheological methods. Specifically, we focus on the elastohydrodynamic coupling between a rigid sphere of radius $R$ and a semi-infinite permeable poroelastic medium. For this purpose, we invoke the linear-response theory introduced by Leroy \& Charlaix~\cite{Leroy_Charlaix_2011}, and widely used in contactless measurements of the mechanical properties of soft surfaces~\cite{Leroy_2012, Villey_2013, guan2017noncontact, guan2017noncontactbis, laine2019micromegascope, Bertin_bubble_2021, zhang2022contactless}. 

\subsection{Soft-lubrication approximation}
The above-mentioned sphere is placed at a distance $D$ from the undeformed gel surface and oscillates vertically with angular frequency $\omega$ and amplitude $h_0$, as schematized in Fig~\ref{fig:schema}(b). The ensemble is fully immersed in a Newtonian fluid (identical to the solvent in the gel here) of dynamic shear viscosity $\eta$ and density $\rho$. We suppose that the sphere-plane distance is small with respect to the sphere radius, and can thus invoke the lubrication approximation~\cite{Reynolds_1886}. The sphere profile can be approximated by a parabola in the lubricated contact region, and the liquid-film thickness profile is thus given by:
\begin{equation}
    \displaystyle h(r,t) \simeq D + h_0 \cos(\omega t) + w(r,t) + \frac{r^2}{2R}.
    \label{eq:gap}
\end{equation}
The Reynolds number is assumed to be small, so that the flow is laminar. Furthermore, we suppose that the typical viscous penetration depth $\sqrt{\eta/(\rho\omega)}$ is large compared to the liquid-gap thickness. Therefore, the flow can be described by the steady Stokes equations with no-slip boundary conditions at both the sphere and gel surfaces. This latter condition is assumed for simplicity since the typical slip length at poroelastic surfaces is comparable to the pore size $\sim\sqrt{k}$~\cite{Beavers_1967, Knox_2017}, which is normally nanometric. 

The liquid-film thickness profile follows the axisymmetric thin-film equation~\cite{GHP_2001}:
\begin{equation}
    \displaystyle \frac{\partial h}{\partial t} = \frac{1}{12 \eta r} \frac{\partial}{\partial r}\left[ rh^3 \frac{\partial p}{\partial r} \right],
    \label{eq:reynolds}
\end{equation}
where $p$ is the excess pressure field in the liquid with respect to the atmospheric pressure. In the lubrication approximation, the viscous shear stresses are negligible compared to the pressure. Therefore, the force balance at the gel surface takes the same form as in section~\ref{sec:sudden-response}, and the surface deformation profile can be computed from Eq.~\eqref{eq:response}.

\subsection{Linear-response theory}
Following Ref.~\cite{Leroy_Charlaix_2011}, we suppose that the oscillation amplitude is much smaller than the liquid-gap thickness. Hence, we invoke the linear-response theory, and write the fields as: 
\begin{equation}
	\label{eq:linear_response_theory}
	w(r,t) = \mathrm{Re}[w^*(r)e^{i\omega t}], \quad p(r,t) = \mathrm{Re}[p^*(r)e^{i\omega t}],
\end{equation}
where $^*$ indicates complex variables, $i^2=-1$, and $\mathrm{Re}$ is the real part. Equation~\eqref{eq:reynolds} is then linearized, giving:
\begin{equation}
	\displaystyle i \omega \Big(h_0 + w^* \Big) = \frac{1}{12 \eta r} \frac{\mathrm{d}}{\mathrm{d} r}\left[r\Big(D + \frac{r^2}{2R} \Big)^3\, \frac{\mathrm{d} p^*}{\mathrm{d} r}\right].
	\label{eq:reynolds_fourier}
\end{equation}
Using the solution derived in section~\ref{sec:generalization}, we can obtain the surface deformation by injecting Eq.~\eqref{eq:linear_response_theory} into Eq.~\eqref{eq:response}. The amplitude of the surface deformation in Hankel space reads:
\begin{equation}
	\hat{w}^*(s) = \hat{p}^*(s)\hat{\mathcal{G}}^*(s) = \frac{\hat{p}^*(s)}{2Gs}\,\frac{1}{1- i\Lambda\frac{\mathcal{D}_\mathrm{pe} s^2}{\omega}\Big(1- \sqrt{1+ \frac{i \omega}{\mathcal{D}_\mathrm{pe} s^2}} \Big)},\\
	\label{eq:Hankel_def}
\end{equation}
with:
\begin{equation}
	\hat{w}^*(s) = \int_0^\infty \mathrm{d}r\,w^*(r)  r\,J_0(sr), \quad \hat{p}^*(s) = \int_0^\infty\mathrm{d}r\, p^*(r)  r\,J_0(sr).
\end{equation}
Note that, in contrast to Sec.II, the $\hat{}$ symbol now only refers to the Hankel transform, since there is no time dependence on the amplitudes and thus no Laplace transform. From Eq.~\eqref{eq:gap}, the contact length $\sqrt{2RD}$ ---\textit{i.e.} the so-called hydrodynamic radius--- sets a typical horizontal length scale. Besides, $h_0$ sets a typical vertical length scale. Thus, we introduce the following dimensionless variables:
\begin{equation}
    \begin{array}{cc}
        \displaystyle \Tilde{r} = \frac{r}{\sqrt{2RD}}, \quad \Tilde{s} = s \sqrt{2RD}, \quad  \Tilde{w}^*(\Tilde{r}) = \frac{w^*(r)}{h_0}. 
    \end{array}
\end{equation}
From the horizontal projection of the Stokes equation, and the incompressibility condition, we find that the typical lubrication pressure scale is $2\eta R \omega h_0/D^2$. Thus, we introduce the following dimensionless pressure field:
\begin{equation}
    \Tilde{p}^*(\Tilde{r}) = \frac{D^2p^*(r)}{2 \omega R \eta h_0}.
\end{equation}
Injecting these new variables in Eq.~\eqref{eq:reynolds}, the dimensionless thin-film equation results:
\begin{equation}
    \displaystyle i(1 + \Tilde{w}^*)=\frac{1}{12 \Tilde{r}} \frac{\mathrm{d}}{\mathrm{d} \Tilde{r}}\left[\Tilde{r} \Big(1+\Tilde{r}^2\Big)^3 \frac{\mathrm{d} \Tilde{p}^*}{\mathrm{d} \Tilde{r}} \right].
    \label{eq:reynolds_nodim_cpx}
\end{equation}
Finally, we introduce two characteristic parameters. First, the critical distance at which the surface deformation and sphere oscillation amplitude are of the same order:
\begin{equation}
     D_\mathrm{c} = 8R\Big( \frac{\eta \omega}{2G}\Big)^{2/3}.
\label{eq:critical_distance}
\end{equation}
Second, the critical poroelastic angular frequency at which solvent typically diffuses over the contact length at the critical distance during one oscillation:
\begin{equation}
\omega_\mathrm{c} = \frac{\mathcal{D}_\mathrm{pe}}{2RD_\mathrm{c}} = \frac{\mathcal{D}_\mathrm{pe}}{16R^2}\Big( \frac{2G}{\eta \omega}\Big)^{2/3}. \\
\label{eq:critical_pulsation}
\end{equation}

\subsection{Deformation profile and normal force}
Using the dimensionless variables and critical parameters above, we can write
Eq.~\eqref{eq:Hankel_def} in dimensionless form, as:
\begin{equation}
	\hat{\tilde{w}}^*(s) = \frac{\hat{\tilde{p}}^*(s)}{8s}\,\frac{(D_\mathrm{c}/D)^{3/2}}{1- i\Lambda\frac{\tilde{s}^2}{\omega D/(\omega_\mathrm{c}D_\mathrm{c})}\Big(1- \sqrt{1+ \frac{i \omega D/(\omega_\mathrm{c}D_\mathrm{c})}{\tilde{s}^2}} \Big)}.
	\label{def_cpx}
\end{equation}
Moreover, the amplitude $F^*$ of the vertical elastohydrodynamic force exerted on the sphere is obtained by integrating the amplitude of the lubrication pressure field over the surface, as:
\begin{equation}
    F^* = 2\pi \int_0^\infty\mathrm{d}r\,r\, p^*(r) =\frac{8\pi \eta\omega h_0 R^2}{D_\mathrm{c}} \tilde{F}^*,\quad \textrm{with}  \quad \tilde{F}^* = \frac{D_\mathrm{c}}{D} \int_0^\infty\mathrm{d}\tilde{r}\,\tilde{r}\, \tilde{p}^*(\tilde{r}),
    \label{force_cpx}
\end{equation}
where we have noted the dimensionless force $\tilde{F}^*$, which depends on the dimensionless parameters, $D/D_\mathrm{c}$, $\omega/\omega_\mathrm{c}$, and $\Lambda$. 
\begin{figure}[t!]
	\centering
	\includegraphics[width = 16 cm]{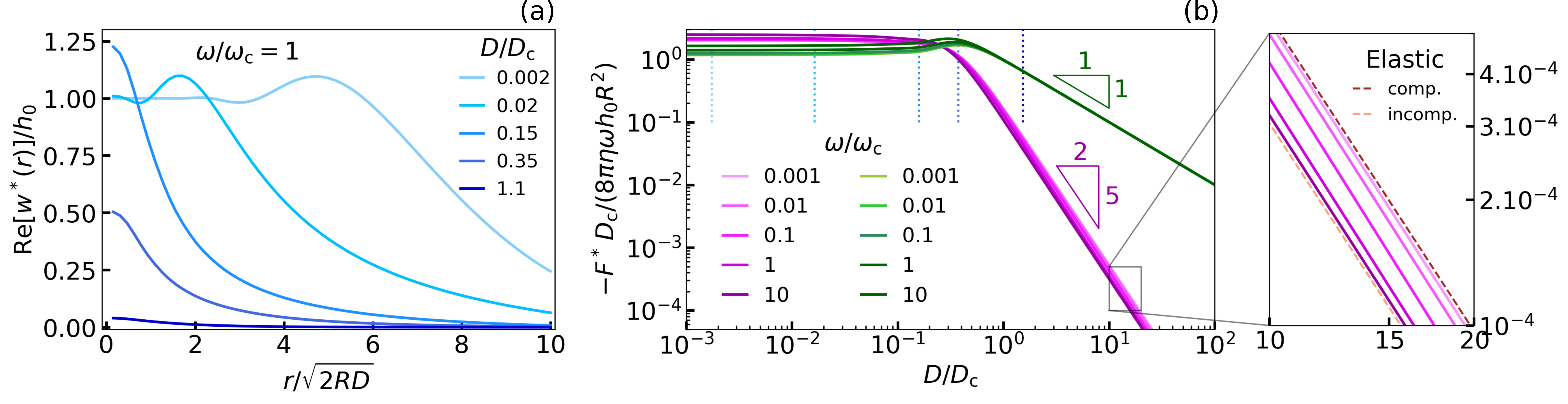}
	\caption{\textbf{Mechanical response of a poroelastic gel in the contactless colloidal-probe configuration.}\textit{(a) Normalized amplitude $w^*/h_0$ of the surface deformation profile as a function of the rescaled radial coordinate $r/{\sqrt{2RD}}$, for various normalized sphere-substrate distances, as computed from Eq.~(\ref{def_cpx}) with $\omega/\omega_{\textrm{c}}=1$ and $\nu=0.1$. (b) Real (pink) and imaginary (green) parts of the normalized force $F^*D_{\textrm{c}}/(8\pi\eta\omega h_0 R^2)$ exerted on the spherical probe as functions of normalized sphere-substrate distance, for various reduced angular frequencies $\omega/\omega_{\mathrm{c}}$, as computed from Eq.~(\ref{force_cpx}) with $\nu=0.1$. The vertical dashed blue lines correspond to the distances at which the surface deformation profiles are plotted in (a). The zoomed inset shows in addition the purely elastic incompressible case (orange) and the purely elastic compressible case (red). }}
	\label{fig:Force_distance}
\end{figure}

Equations (\ref{eq:reynolds_nodim_cpx}) and (\ref{def_cpx}) can be solved numerically, as detailed in Appendix~\ref{Sec:permeable-num}. Examples of the obtained surface deformation field are plotted in Fig.~\ref{fig:Force_distance}(a) for various sphere-substrate distances. In a contactless colloidal-probe rheological experiment, however it is not the deformation amplitude that is typically measured. Rather, the sampled surface slowly approaches the oscillating spherical probe using a piezo stage, with the typical experimental outputs being the measured force amplitude and phase as functions of the sphere-substrate distance. The other parameters are kept constant. From the amplitude and phase, the real and imaginary components of the complex force can be evaluated. These force components can be evaluated theoretically using Eq.~(\ref{force_cpx}), while the amplitude of the pressure field can also be obtained numerically. 

Therefore, in Fig.~\ref{fig:Force_distance}(b), we plot the dimensionless force amplitudes as a function of the dimensionless distance, for various oscillation angular frequencies. Two regimes can be observed. At large distance, \textit{i.e.} $D/D_\mathrm{c}\gg 1$, the surface deformation is small with respect to the oscillation amplitude (see Fig.~\ref{fig:Force_distance}(a)). As a result, the elastohydrodynamic coupling is weak, and the force is dominated by the viscous dissipation in the liquid film. In the far-field asymptotic regime, we recover the classical lubrication Stokes drag $F^* = -6i\pi \eta R^2 h_0\omega/D$, which can be obtained by integrating Eq.~\eqref{eq:reynolds} in the absence of substrate deformation. The real part of the force amplitude in the far field is much smaller than the imaginary part, but it displays a signature of the elasticity of the substrate through an asymptotic decay with the distance as $\sim(D/D_\mathrm{c})^{-5/2}$. The exact prefactor of this scaling law can be obtained by expanding the solution in $(D_\mathrm{c}/D)$~\cite{Leroy_Charlaix_2011, Bertin_bubble_2021}. At small distance, \textit{i.e.} $D/D_\mathrm{c}\ll 1$, the substrate deformation saturates and scales with the oscillation amplitude (see Fig.~\ref{fig:Force_distance}(a)). As a consequence, the real and imaginary parts of the force amplitude saturate as well to values that do not depend on the distance. Besides, at all distances (see \textit{e.g.} inset of Fig.~\ref{fig:Force_distance}(b), at large distance), we recover  the qualitative feature discussed in the previous part: at small frequency, the elastohydrodynamic coupling is similar to the one of a purely elastic compressible layer; conversely, at large frequency, the elastohydrodynamic coupling is similar to the one of a purely elastic incompressible layer. 

Finally, in these rescaled variables, we observe a small influence of the solvent diffusion in the gel on the elastohydrodynamic coupling, despite the diffusion constant being varied over 4 orders of magnitude (via the critical frequency). Therefore, from our model, it appears that contactless colloidal-probe rheological methods in the linear-response regime are not appropriate to robustly measure the effects of the solvent diffusion through the gel network. In contrast, such methods appear to be well suited for measuring the effective shear elastic moduli and Poisson ratios of gels. 

\section{Conclusion}
We  theoretically addressed the mechanical response of a semi-infinite and permeable, linear poroelastic substrate to an external axisymmetric pressure field. The point-force response was first computed. By convolution of the latter to any outer pressure field, the surface deformation profile can be computed. Motivated by the recent development of contactless colloidal-probe rheological experiments on soft and complex materials, we applied our general results to the specific case of a sphere oscillating vertically near a gel, within an outer fluid identical to the solvent present in the polymeric matrix. The complex amplitude of the force exerted on the spherical probe was numerically computed and studied. As a result,  contactless colloidal-probe methods in the linear-response regime appear as good candidates to robustly measure the effective elastic properties of gels and biological membranes, without risks of wear and adhesion. Going beyond linear response, and incorporating large deformations as well as a polymeric description of the gel architecture seem to be the next steps towards modelling further the complex response of gels and measuring their specific poroelastic behaviours. 

\section*{Data accessibility}
The data presented in this article are accessible from the authors upon reasonable request.

\section*{Competing interests}

The authors declare no competing interests. 

\section*{Authors' contributions} 
V.B., J.D.M. and T.S. designed the research. C.K.-M. conducted the research and wrote the first draft of the manuscript. All authors contributed to the analysis of the results and writing of the manuscript.

\section*{Acknowledgments}
The authors thank Elisabeth Charlaix, Abdelhamid Maali, Zaicheng Zhang, and Yvette Tran, for interesting discussions. They also thank the Soft Matter Collaborative Research Unit, Frontier Research Center for Advanced Material and Life Science, Faculty of Advanced Life Science at Hokkaido University, Sapporo, Japan. 

\section*{Funding}
The authors acknowledge financial support from the European Union through the European Research Council under EMetBrown (ERC-CoG-101039103) grant, as well as the Agence Nationale de la Recherche under CoPinS (ANR-19-CE06-0021), EMetBrown (ANR-21-ERCC-0010-01), Softer (ANR-21-CE06-0029) and Fricolas (ANR-21-CE06-0039) grants. The authors also acknowledge financial support from Institut Pierre-Gilles de Gennes (Equipex ANR-10-EQPX-34 and Labex ANR-10-LABX-31). 

\appendix
\section*{Appendix}
\subsection{Long-term asymptotic deformation}
\label{APP:long-time-power}
In order to rationalize the $t^{-1/2}$ power law observed in the inset of Fig~\ref{fig:Point-force_response-real_space}(b), the point-force solution obtained in the main text, Eq.~\ref{eq:green}, can be expanded at small temporal frequency $q$ (or similarly at large time $t$), as:
	\begin{equation}
		\begin{split}
	\hat{w}(s,q)&\approx  \frac{F_0}{4\pi Gsq}\frac{1}{1+\frac{1-2\nu}{1-\nu}\frac{\mathcal{D}_\mathrm{pe}s^2}{q}\Big(1-\left[1+\frac{1}{2}\frac{q}{s^2\mathcal{D}_\mathrm{pe}}-\frac{1}{8}\left(\frac{q}{s^2\mathcal{D}_\mathrm{pe}}\right)^2\right]\Big)} \\
						&= \frac{F_0(1-\nu)}{2\pi Gsq} \frac{1}{1+\frac{(1-2\nu)q}{4\mathcal{D}_\mathrm{pe}s^2}}.
		\end{split}
	\end{equation}
Taking the inverse Laplace transform, we get: 
	\begin{equation}
	\hat{w}(s,t) \approx \frac{F_0(1-\nu)}{2\pi Gs} \left[1-\exp\left(-\frac{4\mathcal{D}_\mathrm{pe}t s^2}{1-2\nu}\right)\right].
	\end{equation}
	Finally taking the inverse Hankel transform, we get: 
	\begin{equation}
				\begin{split}
		w(r,t) &= \frac{F_0(1-\nu)}{2\pi G r} \left[1 - \frac{r}{\sqrt{\frac{\mathcal{D}_\mathrm{pe}t}{\pi(1-2\nu)}}}I_0\left(-\frac{(1-2\nu)r^2}{32\mathcal{D}_\mathrm{pe}t}\right)\exp\left(-\frac{(1-2\nu)r^2}{32\mathcal{D}_\mathrm{pe}t}\right)\right]\\
				 &\simeq \frac{F_0(1-\nu)}{2\pi G r} - \frac{F_0(1-\nu)\sqrt{(1-2\nu)\pi}}{2\pi G \sqrt{16\mathcal{D}_\mathrm{pe}t}},
				\end{split}
	\end{equation}
where $I_0$ is a modified Bessel function of the first kind, of order 0, and the last expansion is obtained by taking the long-time limit. The first term of the right-hand side gives the purely elastic and compressible response of the material at long time. The second term corresponds to the long-term correction to the latter, as plotted in Fig.~\ref{fig:Point-force_response-real_space}(b). The decay does not depend on $r$ and scales as $\sim1/\sqrt{\mathcal{D}_\mathrm{pe}t}$, as recovered through the asymptotic $-1/2$ exponent in the inset of Fig.~\ref{fig:Point-force_response-real_space}(b).

\subsection{Point-force solution for an impermeable gel and comparison to the permeable case}
\label{app:impermeable_comp}
Here, we discuss the effect of surface permeability on the point-force response. In the model presented in the main text, we assumed that the gel is in contact with a bath of its own solvent, which fixes the chemical potential of the solvent at the surface to its reference value, at all times. If the gel is in contact with an other kind of medium (\textit{e.g.} air, solid surface, or immiscible liquid), the description should be modified, and incorporate drying in particular. If one considers time scales smaller than the drying time, we can neglect the solvent exchange at the interface. In that case, we suppose the solvent flux vanishes at the surface, allowing us to impose the \textit{impermeable} boundary condition at $z=0$:
\begin{equation}
	\mathbf{J}\cdot \mathbf{n} = \frac{\partial \mu}{\partial z} = 0.
\end{equation}
We can use the same method as the one used in the main text, in order to derive the point-force response:
\begin{equation}
	\displaystyle \hat{w}^{\text{imper}}(s,q)= - \frac{1}{2Gsq}\frac{1}{1+\Lambda\frac{\mathcal{D}_\mathrm{pe}s^2}{q}\Big(\frac{1}{\sqrt{1+\frac{q}{s^2\mathcal{D}_\mathrm{pe}}}}-1\Big)},
	\label{green_imper}
\end{equation}
where the superscript ``imper'' stands for the impermeable condition. We thus recover the solution derived in Ref.~\cite{Zhao_2018}. The diffusive-like self-similarity discussed in the main text, holds as well here. 

\begin{figure}[t!]
	\centering
	\includegraphics[width = 16 cm]{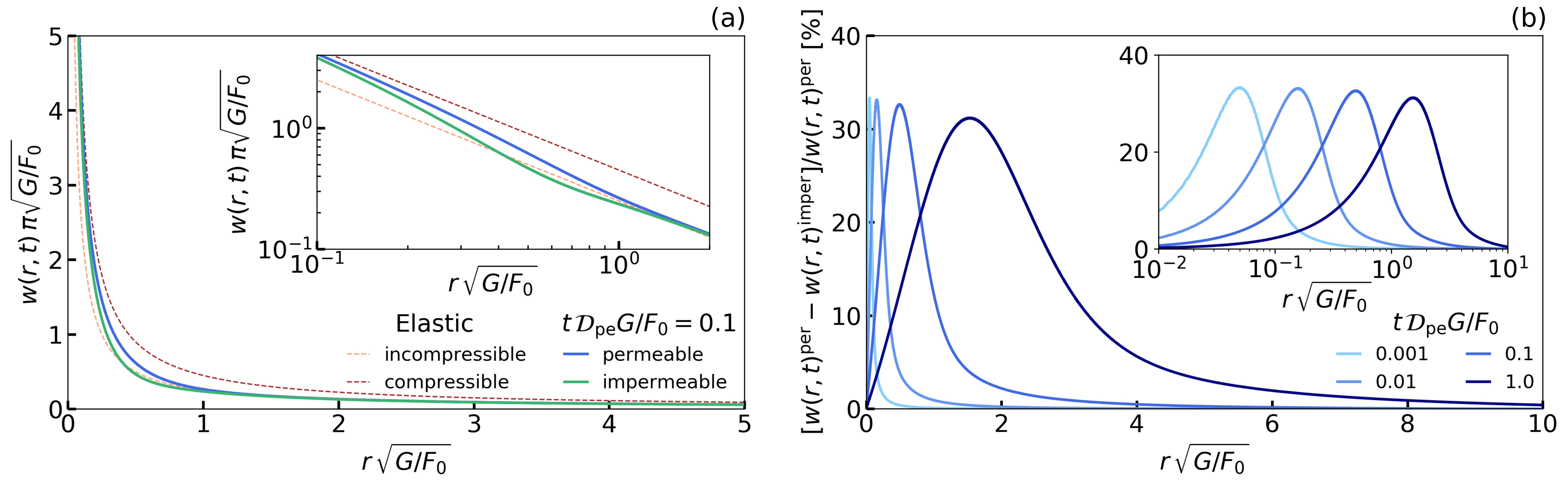}
	\caption{\textbf{Comparison between the point-force solutions of the permeable and impermeable cases.} \textit{(a) Normalized surface deformation as a function of the radial coordinate at the noted time, using $\nu=0.1$. Shown are the point-force solutions for permeable and impermeable boundary conditions, computed from the inverse transforms of Eqs.~\eqref{eq:green} and~\eqref{green_imper}. The orange and red dashed lines correspond to the inverse transforms of Eqs.~\eqref{eqshort} and~\eqref{eqlong}. The inset shows the same data on logarithmic scales. (b) Relative difference between the surface deformation in the permeable and impermeable cases, as a function of the radial coordinate, for the noted dimensionless times. The inset shows the same data in semi-logarithmic scale.}}
	\label{per_imper}
\end{figure}

The point-force responses with impermeable and permeable boundary conditions are compared in Fig.\ref{per_imper}(a). Both solutions appear to be qualitatively similar, and display the same short-term and long-term behaviors. Nevertheless, at a given time $t=0.1F_0/(G\mathcal{D}_\mathrm{pe})$, we observe that the surface deformation in the impermeable case is smaller than the one in the permeable case. Also, the permeable solution relaxes faster than the impermeable one towards the long-time purely elastic compressible limit. Quantitatively, for a Poisson ratio $\nu=0.1$, we observe a relative difference between the two solutions up to $35\%$, where the maximum difference is located at a radial position $\sim \sqrt{\mathcal{D}_\mathrm{pe}t}$, as shown in Fig.~\ref{per_imper} (b). 

\subsection{Numerical computation of the normal force}
\label{Sec:permeable-num}
Here, we detail how Eqs.~(\ref{eq:reynolds_nodim_cpx}), (\ref{def_cpx}) and (\ref{force_cpx}) are numerically solved. All variables are dimensionless in this section for simplicity and the $\Tilde{}$ symbol is omitted. The normalized deformation in reciprocal spaces reads according to Eq.~(\ref{def_cpx}):
\begin{equation}
	\displaystyle 	\hat{w}^*(s) = \frac{\hat{p}^*(s) }{8}\,\Big(\frac{D_\mathrm{c}}{D}\Big)^{3/2}\,\hat{\mathcal{G}}^*(s)\, \quad \text{with} \quad  \hat{\mathcal{G}}^*(s) = \frac{1}{s}\,\frac{1}{1- i\Lambda\frac{s^2}{\omega D/(\omega_\mathrm{c}D_\mathrm{c})}\Big(1- \sqrt{1+ \frac{i \omega D/(\omega_\mathrm{c}D_\mathrm{c})}{s^2}} \Big)}.
	\label{def_fourier_hankel}
\end{equation}
By integrating Eq.~(\ref{eq:reynolds_nodim_cpx}) between $0$ and $r$, and invoking Eq.~(\ref{def_fourier_hankel}), we obtain the following equation for the pressure:
\begin{equation}
	\displaystyle \frac{\mathrm{d}  p^*}{\mathrm{d} r} = \frac{6ir}{(1+r^2)^3} +\frac{3i}{2(1+r^2)^3} \Big(\frac{D_\mathrm{c}}{D}\Big)^{3/2} \int_0^\infty \text{d}s\, J_1(sr)\hat{\mathcal{G}}^*(s)\hat{p}^*(s) .
\end{equation}
Finally, by performing a first-order Hankel transform of the latter equation, we obtain the Fredholm equation of the second kind:
\begin{equation}
	\displaystyle \hat{p}^*(s) = -\frac{3is}{4}K_1(s) - \frac{3i}{2} \Big(\frac{D_\mathrm{c}}{D}\Big)^{3/2}\int_0^\infty \text{d}k \hat{\mathcal{G}}^*(k)\hat{p}^*(k) \int_0^\infty \text{d}r \frac{rJ_1(kr)J_1(sr)}{s(1+r^2)^3},
	\label{algebra}
\end{equation}
where $K_n$ is the modified Bessel function of the second kind with index $n$. The kernel of the Fredholm equation has an analytical solution~\cite{guan2017noncontactbis}, given by:
\begin{equation}
	\begin{split} 
		\int_0^\infty \mathrm{d}r \, \frac{rJ_1(kr)J_1(sr)}{(1+r^2)^3} &= \frac{k^2+s^2}{8} I_1(s) K_1\left(k\right)-sk\frac{I_2(s) K_2\left(k\right)}{4} \quad \mathrm{for} \, \, s<k,\\ 
		&= \frac{k^2+s^2}{8} I_1(k) K_1\left(s\right)-sk\frac{I_2(k) K_2\left(s\right)}{4} \quad \mathrm{for} \, \, k<s,
	\end{split}
\end{equation}
where $I_n$ is the modified Bessel function of the first kind with index $n$. Integrals are numerically evaluated with the Gauss-Legendre-quadrature method. The discretized version of Eq.~(\ref{algebra}) is a linear algebraic problem and can be numerically solved. Finally, the dimensionless force can be computed as in Eq.~(\ref{force_cpx}) by:
\begin{equation}
	\displaystyle F^* = \frac{D_\mathrm{c}}{D}\,\hat{p}^*(s=0).
\end{equation}

\bibliographystyle{ieeetr}
\bibliography{biblio.bib}

\begin{thebibliography}{10}

\bibitem{Andreotti_2016}
B.~Andreotti, O.~B{\"a}umchen, F.~Boulogne, K.~E. Daniels, E.~R. Dufresne,
  H.~Perrin, T.~Salez, J.~H. Snoeijer, and R.~W. Style, ``Solid capillarity:
  when and how does surface tension deform soft solids?,'' {\em Soft Matter},
  vol.~12, no.~12, pp.~2993--2996, 2016.

\bibitem{deramo_2018}
L.~d'Eramo, B.~Chollet, M.~Leman, E.~Martwong, M.~Li, H.~Geisler, J.~Dupire,
  M.~Kerdraon, C.~Vergne, F.~Monti, {\em et~al.}, ``Microfluidic actuators
  based on temperature-responsive hydrogels,'' {\em Microsystems \&
  Nanoengineering}, vol.~4, no.~1, pp.~1--7, 2018.

\bibitem{Flory_1953}
P.~J. Flory, {\em Principles of polymer chemistry}.
\newblock Cornell university press, 1953.

\bibitem{Li_Tran_2015}
M.~Li, B.~Bresson, F.~Cousin, C.~Fretigny, and Y.~Tran, ``Submicrometric films
  of surface-attached polymer network with temperature-responsive properties,''
  {\em Langmuir}, vol.~31, no.~42, pp.~11516--11524, 2015.

\bibitem{Biot_1941}
M.~A. Biot, ``General theory of three-dimensional consolidation,'' {\em Journal
  of applied physics}, vol.~12, no.~2, pp.~155--164, 1941.

\bibitem{hong2008theory}
W.~Hong, X.~Zhao, J.~Zhou, and Z.~Suo, ``A theory of coupled diffusion and
  large deformation in polymeric gels,'' {\em Journal of the Mechanics and
  Physics of Solids}, vol.~56, no.~5, pp.~1779--1793, 2008.

\bibitem{bouklas2012swelling}
N.~Bouklas and R.~Huang, ``Swelling kinetics of polymer gels: comparison of
  linear and nonlinear theories,'' {\em Soft Matter}, vol.~8, no.~31,
  pp.~8194--8203, 2012.

\bibitem{hu2012viscoelasticity}
Y.~Hu and Z.~Suo, ``Viscoelasticity and poroelasticity in elastomeric gels,''
  {\em Acta Mechanica Solida Sinica}, vol.~25, no.~5, pp.~441--458, 2012.

\bibitem{Zhao_2018}
M.~Zhao, J.~Dervaux, T.~Narita, F.~Lequeux, L.~Limat, and M.~Roch{\'e},
  ``Geometrical control of dissipation during the spreading of liquids on soft
  solids,'' {\em Proceedings of the National Academy of Sciences}, vol.~115,
  no.~8, pp.~1748--1753, 2018.

\bibitem{liu2020coupled}
Z.~Liu, N.~Bouklas, and C.-Y. Hui, ``Coupled flow and deformation fields due to
  a line load on a poroelastic half space: effect of surface stress and surface
  bending,'' {\em Proceedings of the Royal Society A}, vol.~476, no.~2233,
  p.~20190761, 2020.

\bibitem{ang2020effect}
I.~Ang, Z.~Liu, J.~Kim, C.-Y. Hui, and N.~Bouklas, ``Effect of
  elastocapillarity on the swelling kinetics of hydrogels,'' {\em Journal of
  the Mechanics and Physics of Solids}, vol.~145, p.~104132, 2020.

\bibitem{Dervaux_2012}
J.~Dervaux and M.~B. Amar, ``Mechanical instabilities of gels,'' {\em Annu.
  Rev. Condens. Matter Phys.}, vol.~3, no.~1, pp.~311--332, 2012.

\bibitem{johnson_1987}
K.~L. Johnson, {\em Contact mechanics}.
\newblock Cambridge university press, 1987.

\bibitem{hui2006contact}
C.-Y. Hui, Y.~Y. Lin, F.-C. Chuang, K.~R. Shull, and W.-C. Lin, ``A contact
  mechanics method for characterizing the elastic properties and permeability
  of gels,'' {\em Journal of Polymer Science Part B: Polymer Physics}, vol.~44,
  no.~2, pp.~359--370, 2006.

\bibitem{Hu_2010}
Y.~Hu, X.~Zhao, J.~J. Vlassak, and Z.~Suo, ``Using indentation to characterize
  the poroelasticity of gels,'' {\em Applied Physics Letters}, vol.~96, no.~12,
  p.~121904, 2010.

\bibitem{hu2011indentation}
Y.~Hu, X.~Chen, G.~M. Whitesides, J.~J. Vlassak, and Z.~Suo, ``Indentation of
  polydimethylsiloxane submerged in organic solvents,'' {\em Journal of
  Materials Research}, vol.~26, no.~6, pp.~785--795, 2011.

\bibitem{Delavoipiere_2016}
J.~Delavoipiere, Y.~Tran, E.~Verneuil, and A.~Chateauminois, ``Poroelastic
  indentation of mechanically confined hydrogel layers,'' {\em Soft Matter},
  vol.~12, no.~38, pp.~8049--8058, 2016.

\bibitem{Degen_2020}
G.~D. Degen, Y.-T. Chen, A.~L. Chau, L.~K. M{\aa}nsson, and A.~A. Pitenis,
  ``Poroelasticity of highly confined hydrogel films measured with a surface
  forces apparatus,'' {\em Soft Matter}, vol.~16, no.~35, pp.~8096--8100, 2020.

\bibitem{Reynolds_1886}
O.~Reynolds, ``Iv. on the theory of lubrication and its application to mr.
  beauchamp tower’s experiments, including an experimental determination of
  the viscosity of olive oil,'' {\em Philosophical transactions of the Royal
  Society of London}, no.~177, pp.~157--234, 1886.

\bibitem{Vakarelski_2010}
I.~U. Vakarelski, R.~Manica, X.~Tang, S.~J. O’Shea, G.~W. Stevens,
  F.~Grieser, R.~R. Dagastine, and D.~Y. Chan, ``Dynamic interactions between
  microbubbles in water,'' {\em Proceedings of the National Academy of
  Sciences}, vol.~107, no.~25, pp.~11177--11182, 2010.

\bibitem{Leroy_Charlaix_2011}
S.~Leroy and E.~Charlaix, ``Hydrodynamic interactions for the measurement of
  thin film elastic properties,'' {\em Journal of Fluid Mechanics}, vol.~674,
  pp.~389--407, 2011.

\bibitem{dowson2014elasto}
D.~Dowson and G.~R. Higginson, {\em Elasto-hydrodynamic lubrication:
  international series on materials science and technology}.
\newblock Elsevier, 2014.

\bibitem{Kaveh_2014}
F.~Kaveh, J.~Ally, M.~Kappl, and H.-J. Butt, ``Hydrodynamic force between a
  sphere and a soft, elastic surface,'' {\em Langmuir}, vol.~30, no.~39,
  pp.~11619--11624, 2014.

\bibitem{Wang_2015}
Y.~Wang, C.~Dhong, and J.~Frechette, ``Out-of-contact elastohydrodynamic
  deformation due to lubrication forces,'' {\em Physical review letters},
  vol.~115, no.~24, p.~248302, 2015.

\bibitem{Wang_2017_SM}
Y.~Wang, M.~R. Tan, and J.~Frechette, ``Elastic deformation of soft coatings
  due to lubrication forces,'' {\em Soft Matter}, vol.~13, no.~38,
  pp.~6718--6729, 2017.

\bibitem{Wang_2017_COCI}
Y.~Wang, G.~A. Pilkington, C.~Dhong, and J.~Frechette, ``Elastic deformation
  during dynamic force measurements in viscous fluids,'' {\em Current opinion
  in colloid \& interface science}, vol.~27, pp.~43--49, 2017.

\bibitem{Wang_2018}
Y.~Wang and J.~Frechette, ``Morphology of soft and rough contact via fluid
  drainage,'' {\em Soft Matter}, vol.~14, no.~37, pp.~7605--7614, 2018.

\bibitem{Karan_2020}
P.~Karan, S.~S. Das, R.~Mukherjee, J.~Chakraborty, and S.~Chakraborty, ``Flow
  and deformation characteristics of a flexible microfluidic channel with axial
  gradients in wall elasticity,'' {\em Soft Matter}, vol.~16, no.~24,
  pp.~5777--5786, 2020.

\bibitem{Leroy_2012}
S.~Leroy, A.~Steinberger, C.~Cottin-Bizonne, F.~Restagno, L.~L{\'e}ger, and
  E.~Charlaix, ``Hydrodynamic interaction between a spherical particle and an
  elastic surface: a gentle probe for soft thin films,'' {\em Physical review
  letters}, vol.~108, no.~26, p.~264501, 2012.

\bibitem{guan2017noncontactbis}
D.~Guan, C.~Barraud, E.~Charlaix, and P.~Tong, ``Noncontact viscoelastic
  measurement of polymer thin films in a liquid medium using long-needle atomic
  force microscopy,'' {\em Langmuir}, vol.~33, no.~6, pp.~1385--1390, 2017.

\bibitem{zhang2022contactless}
Z.~Zhang, M.~Arshad, V.~Bertin, S.~Almohamad, E.~Rapha{\"e}l, T.~Salez, and
  A.~Maali, ``Contactless rheology of soft gels over a broad frequency range,''
  {\em arXiv preprint arXiv:2202.04386}, 2022.

\bibitem{Villey_2013}
R.~Villey, E.~Martinot, C.~Cottin-Bizonne, M.~Phaner-Goutorbe, L.~L{\'e}ger,
  F.~Restagno, and E.~Charlaix, ``Effect of surface elasticity on the rheology
  of nanometric liquids,'' {\em Physical review letters}, vol.~111, no.~21,
  p.~215701, 2013.

\bibitem{guan2017noncontact}
D.~Guan, E.~Charlaix, R.~Z. Qi, and P.~Tong, ``Noncontact viscoelastic imaging
  of living cells using a long-needle atomic force microscope with
  dual-frequency modulation,'' {\em Physical Review Applied}, vol.~8, no.~4,
  p.~044010, 2017.

\bibitem{Wang_Maali_2018}
Y.~Wang, B.~Zeng, H.~T. Alem, Z.~Zhang, E.~Charlaix, and A.~Maali,
  ``Viscocapillary response of gas bubbles probed by thermal noise atomic force
  measurement,'' {\em Langmuir}, vol.~34, no.~4, pp.~1371--1375, 2018.

\bibitem{Bertin_bubble_2021}
V.~Bertin, Z.~Zhang, R.~Boisgard, C.~Grauby-Heywang, E.~Raphael, T.~Salez, and
  A.~Maali, ``Contactless rheology of finite-size air-water interfaces,'' {\em
  Physical Review Research}, vol.~3, no.~3, p.~L032007, 2021.

\bibitem{Cappella_Dietler_1999}
B.~Cappella and G.~Dietler, ``Force-distance curves by atomic force
  microscopy,'' {\em Surface science reports}, vol.~34, no.~1-3, pp.~1--104,
  1999.

\bibitem{Butt_2005}
H.-J. Butt, B.~Cappella, and M.~Kappl, ``Force measurements with the atomic
  force microscope: Technique, interpretation and applications,'' {\em Surface
  science reports}, vol.~59, no.~1-6, pp.~1--152, 2005.

\bibitem{Israel_1976}
J.~N. Israelachvili and G.~Adams, ``Direct measurement of long range forces
  between two mica surfaces in aqueous kno3 solutions,'' {\em Nature},
  vol.~262, no.~5571, pp.~774--776, 1976.

\bibitem{Israel_2010}
J.~Israelachvili, Y.~Min, M.~Akbulut, A.~Alig, G.~Carver, W.~Greene,
  K.~Kristiansen, E.~Meyer, N.~Pesika, K.~Rosenberg, {\em et~al.}, ``Recent
  advances in the surface forces apparatus (sfa) technique,'' {\em Reports on
  Progress in Physics}, vol.~73, no.~3, p.~036601, 2010.

\bibitem{Kristiansen_2019}
K.~Kristiansen, S.~H. Donaldson~Jr, Z.~J. Berkson, J.~Scott, R.~Su, X.~Banquy,
  D.~W. Lee, H.~B. De~Aguiar, J.~D. McGraw, G.~D. Degen, {\em et~al.},
  ``Multimodal miniature surface forces apparatus ($\mu$sfa) for interfacial
  science measurements,'' {\em Langmuir}, vol.~35, no.~48, pp.~15500--15514,
  2019.

\bibitem{laine2019micromegascope}
A.~Lain{\'e}, L.~Jubin, L.~Canale, L.~Bocquet, A.~Siria, S.~H. Donaldson, and
  A.~Nigu{\`e}s, ``Micromegascope based dynamic surface force apparatus,'' {\em
  Nanotechnology}, vol.~30, no.~19, p.~195502, 2019.

\bibitem{Sekimoto_1993}
K.~Sekimoto and L.~Leibler, ``A mechanism for shear thickening of
  polymer-bearing surfaces: elasto-hydrodynamic coupling,'' {\em EPL
  (Europhysics Letters)}, vol.~23, no.~2, p.~113, 1993.

\bibitem{Beaucourt_2004}
J.~Beaucourt, T.~Biben, and C.~Misbah, ``Optimal lift force on vesicles near a
  compressible substrate,'' {\em EPL (Europhysics Letters)}, vol.~67, no.~4,
  p.~676, 2004.

\bibitem{Skotheim_Maha_2005}
J.~Skotheim and L.~Mahadevan, ``Soft lubrication: The elastohydrodynamics of
  nonconforming and conforming contacts,'' {\em Physics of Fluids}, vol.~17,
  no.~9, p.~092101, 2005.

\bibitem{Urzay_2007}
J.~Urzay, S.~G. Llewellyn~Smith, and B.~J. Glover, ``The elastohydrodynamic
  force on a sphere near a soft wall,'' {\em Physics of Fluids}, vol.~19,
  no.~10, p.~103106, 2007.

\bibitem{Snoeijer_2013}
J.~H. Snoeijer, J.~Eggers, and C.~H. Venner, ``Similarity theory of lubricated
  hertzian contacts,'' {\em Physics of fluids}, vol.~25, no.~10, p.~101705,
  2013.

\bibitem{Salez_Maha_2015}
T.~Salez and L.~Mahadevan, ``Elastohydrodynamics of a sliding, spinning and
  sedimenting cylinder near a soft wall,'' {\em Journal of Fluid Mechanics},
  vol.~779, pp.~181--196, 2015.

\bibitem{Vialar_2019}
P.~Vialar, P.~Merzeau, S.~Giasson, and C.~Drummond, ``Compliant surfaces under
  shear: elastohydrodynamic lift force,'' {\em Langmuir}, vol.~35, no.~48,
  pp.~15605--15613, 2019.

\bibitem{Zhang_2020}
Z.~Zhang, V.~Bertin, M.~Arshad, E.~Raphael, T.~Salez, and A.~Maali, ``Direct
  measurement of the elastohydrodynamic lift force at the nanoscale,'' {\em
  Physical review letters}, vol.~124, no.~5, p.~054502, 2020.

\bibitem{Bertin_lift_2022}
V.~Bertin, Y.~Amarouchene, E.~Raphael, and T.~Salez, ``Soft-lubrication
  interactions between a rigid sphere and an elastic wall,'' {\em Journal of
  Fluid Mechanics}, vol.~933, 2022.

\bibitem{Bureau_2022}
L.~Bureau, G.~Coupier, and T.~Salez, ``Lift at zero reynolds number,'' {\em
  arXiv preprint arXiv:2207.04538}, 2022.

\bibitem{Bouchet_2015}
A.-S. Bouchet, C.~Cazeneuve, N.~Baghdadli, G.~S. Luengo, and C.~Drummond,
  ``Experimental study and modeling of boundary lubricant polyelectrolyte
  films,'' {\em Macromolecules}, vol.~48, no.~7, pp.~2244--2253, 2015.

\bibitem{Saintyves_2016}
B.~Saintyves, T.~Jules, T.~Salez, and L.~Mahadevan, ``Self-sustained lift and
  low friction via soft lubrication,'' {\em Proceedings of the National Academy
  of Sciences}, vol.~113, no.~21, pp.~5847--5849, 2016.

\bibitem{Davies_2018}
H.~S. Davies, D.~D{\'e}barre, N.~El~Amri, C.~Verdier, R.~P. Richter, and
  L.~Bureau, ``Elastohydrodynamic lift at a soft wall,'' {\em Physical review
  letters}, vol.~120, no.~19, p.~198001, 2018.

\bibitem{Rallabandi_2018}
B.~Rallabandi, N.~Oppenheimer, M.~Y. Ben~Zion, and H.~A. Stone,
  ``Membrane-induced hydroelastic migration of a particle surfing its own
  wave,'' {\em Nature Physics}, vol.~14, no.~12, pp.~1211--1215, 2018.

\bibitem{Hou_1992}
J.~Hou, V.~C. Mow, W.~Lai, and M.~Holmes, ``An analysis of the squeeze-film
  lubrication mechanism for articular cartilage,'' {\em Journal of
  biomechanics}, vol.~25, no.~3, pp.~247--259, 1992.

\bibitem{Jahn_Klein_2018}
S.~Jahn, J.~Seror, and J.~Klein, ``Lubrication of articular cartilage,'' {\em
  Annual review of biomedical engineering}, vol.~18, pp.~235--258, 2016.

\bibitem{Cher_2008}
I.~Cher, ``A new look at lubrication of the ocular surface: fluid mechanics
  behind the blinking eyelids,'' {\em The ocular surface}, vol.~6, no.~2,
  pp.~79--86, 2008.

\bibitem{pandey2016lubrication}
A.~Pandey, S.~Karpitschka, C.~H. Venner, and J.~H. Snoeijer, ``Lubrication of
  soft viscoelastic solids,'' {\em Journal of fluid mechanics}, vol.~799,
  pp.~433--447, 2016.

\bibitem{Kargar_2021}
A.~Kargar-Estahbanati and B.~Rallabandi, ``Lift forces on three-dimensional
  elastic and viscoelastic lubricated contacts,'' {\em Physical Review Fluids},
  vol.~6, no.~3, p.~034003, 2021.

\bibitem{hui2021friction}
C.-Y. Hui, H.~Wu, A.~Jagota, and C.~Khripin, ``Friction force during lubricated
  steady sliding of a rigid cylinder on a viscoelastic substrate,'' {\em
  Tribology Letters}, vol.~69, no.~2, pp.~1--17, 2021.

\bibitem{Delavoipiere_2018}
J.~Delavoipi{\`e}re, Y.~Tran, E.~Verneuil, B.~Heurtefeu, C.~Y. Hui, and
  A.~Chateauminois, ``Friction of poroelastic contacts with thin hydrogel
  films,'' {\em Langmuir}, vol.~34, no.~33, pp.~9617--9626, 2018.

\bibitem{Ciapa_2020}
L.~Ciapa, J.~Delavoipi{\`e}re, Y.~Tran, E.~Verneuil, and A.~Chateauminois,
  ``Transient sliding of thin hydrogel films: the role of poroelasticity,''
  {\em Soft Matter}, vol.~16, no.~28, pp.~6539--6548, 2020.

\bibitem{cuccia2020pore}
N.~L. Cuccia, S.~Pothineni, B.~Wu, J.~M{\'e}ndez~Harper, and J.~C. Burton,
  ``Pore-size dependence and slow relaxation of hydrogel friction on smooth
  surfaces,'' {\em Proceedings of the National Academy of Sciences}, vol.~117,
  no.~21, pp.~11247--11256, 2020.

\bibitem{Alaa_2022}
M.~A. Eddine, S.~Belbekhouche, S.~de~Chateauneuf-Randon, T.~Salez,
  A.~Kovalenko, B.~Bresson, and C.~Monteux, ``Large and nonlinear permeability
  amplification with polymeric additives in hydrogel membranes,'' {\em
  Macromolecules}, 2022.

\bibitem{Beavers_1967}
G.~S. Beavers and D.~D. Joseph, ``Boundary conditions at a naturally permeable
  wall,'' {\em Journal of fluid mechanics}, vol.~30, no.~1, pp.~197--207, 1967.

\bibitem{Meeker_2004}
S.~P. Meeker, R.~T. Bonnecaze, and M.~Cloitre, ``Slip and flow in pastes of
  soft particles: Direct observation and rheology,'' {\em Journal of Rheology},
  vol.~48, no.~6, pp.~1295--1320, 2004.

\bibitem{Knox_2017}
D.~Knox, B.~Duffy, S.~McKee, and S.~Wilson, ``Squeeze-film flow between a
  curved impermeable bearing and a flat porous bed,'' {\em Physics of Fluids},
  vol.~29, no.~2, p.~023101, 2017.

\bibitem{PVdV_2022}
P.~Van~de Velde, J.~Dervaux, S.~Proti{\`e}re, and C.~Duprat, ``Spontaneous
  localized fluid release on swelling fibres,'' {\em Soft Matter}, vol.~18,
  no.~24, pp.~4565--4571, 2022.

\bibitem{Hughes_1979}
B.~Hughes and L.~White, ``‘soft’contact problems in linear elasticity,''
  {\em The Quarterly Journal of Mechanics and Applied Mathematics}, vol.~32,
  no.~4, pp.~445--471, 1979.

\bibitem{Essink2020}
M.~H. Essink, A.~Pandey, S.~Karpitschka, C.~H. Venner, and J.~H. Snoeijer,
  ``Regimes of soft lubrication,'' {\em Journal of fluid mechanics}, vol.~915,
  2021.

\bibitem{MacNamee_Gibson_1}
J.~McNAMEE and R.~Gibson, ``Plane strain and axially symmetric problems of the
  consolidation of a semi-infinite clay stratum,'' {\em The Quarterly Journal
  of Mechanics and Applied Mathematics}, vol.~13, no.~2, pp.~210--227, 1960.

\bibitem{MacNamee_Gibson_2}
J.~McNamee and R.~Gibson, ``Displacement functions and linear transforms
  applied to diffusion through porous elastic media,'' {\em The Quarterly
  Journal of Mechanics and Applied Mathematics}, vol.~13, no.~1, pp.~98--111,
  1960.

\bibitem{Gibson_1970}
R.~Gibson, R.~Schiffman, and S.~Pu, ``Plane strain and axially symmetric
  consolidation of a clay layer on a smooth impervious base,'' {\em The
  Quarterly Journal of Mechanics and Applied Mathematics}, vol.~23, no.~4,
  pp.~505--520, 1970.

\bibitem{Hariprasad_2012}
D.~S. Hariprasad and T.~W. Secomb, ``Motion of red blood cells near microvessel
  walls: effects of a porous wall layer,'' {\em Journal of fluid mechanics},
  vol.~705, pp.~195--212, 2012.

\bibitem{Abate_Whitt_2006}
J.~Abate and W.~Whitt, ``A unified framework for numerically inverting laplace
  transforms,'' {\em INFORMS Journal on Computing}, vol.~18, no.~4,
  pp.~408--421, 2006.

\bibitem{Xu_2020}
Q.~Xu, L.~A. Wilen, K.~E. Jensen, R.~W. Style, and E.~R. Dufresne,
  ``Viscoelastic and poroelastic relaxations of soft solid surfaces,'' {\em
  Physical Review Letters}, vol.~125, no.~23, p.~238002, 2020.

\bibitem{GHP_2001}
E.~Guyon, J.-P. Hulin, L.~Petit, and P.~G. de~Gennes, {\em Hydrodynamique
  physique}.
\newblock EDP sciences Les Ulis, 2001.

\end{thebibliography}

\end{document}